\documentclass{cpbtex}
\begin{document}
\begin{CJK*}{GBK}{song}

\title{The global monopole spacetime and its
topological charge \thanks{Project supported by the National Natural Science Foundation of China (Grant No.~11273009 and 11303006).}}


\author{Hongwei Tan$^{1}$, \ Jinbo Yang $^{1,2}$, \ Jingyi Zhang $^{1}$\thanks{Jingyi Zhang. E-mail:zzhangjy@gzhu.edu.cn}, \ and \\ Tangmei He $^{1}$\\
$^{1}${Center for Astrophysics, Guangzhou University, Guangzhou 510006}\\  
$^{2}${co-first author}}   


\date{\today}
\maketitle

\begin{abstract}
In this paper, we show that the global monopole spacetime is one of the exact solutions of Einstein equations  by means of the method treating the matter field as a non-linear sigma model, without the weak field approximation applied  in the original derivation by  Barriola and Vilenkin in 1989. Furthermore, we find the physical
origin of the topological charge in the global monopole spacetime. Finally, we generalize the proposal which generates spacetime from thermodynamical laws to the case that spacetime with global monopole charge.
\end{abstract}

\maketitle
\section{Introduction}
The global monopole charge resulted from the breaking
of global O(3) symmetry belongs to those eccentric objects like cosmic strings, domain walls and etc, and playing an important role
in modern physics,
 see \ucite{vilenkin}. Generally speaking, the global monopole charge is one kind of topological defect which can be produced by the phase transition in the early universe.
There were a series of important investigations about this kind of topic before. For instance,  Harari and Loust\`{o} found that the gravitational potential of such topological defect is negative \ucite{Harari};
 the same topic in the dS/AdS spacetime was explored in Ref. \cite{Li} and the scholars found that the gravitational potential can be either  repulsive or attractive, depending on the  value of the cosmological
constant. In addition, a new class of cold stars called D-stars (defect stars) were proposed in Ref. \cite{Li2,Li3,Li4} in which the authors argued that the theory has monopole solutions in the case that
the matter field disappears. Besides, the global monopole charge has also been considered in cosmology, see Ref. \cite{vilenkin2,Basu,Basu2,Li5} for more details.
Moreover, the global monopole spacetime was first found by Barriola and Vilenkin in 1989 \ucite{Barriola} in which this category of spacetime was derived by applying the  weak field approximation.

In Ref. \cite{tian}, the general bulk/boundary correspondence, a generalized version of the AdS/CFT correspondence, was proposed.     In order to find the evidence  to support their proposal, the authors
put forward   the thermodynamics on the bulk where the topological charge is constructed. However, it should be noted that  the authors' proposal of this new physical concept (namely the topological charge) was based on some mathematical
tricks and the physical origin of the topological charge was missed in this proposal.

In this paper we prove that the global monopole
spacetime is an exact solution of the  Einstein equations if the matter field is applied as the free nonlinear sigma model which has been introduced in textbook\ucite{peskin}. Additionally, we show that our work can be applied to higher
dimensional spacetime as well as another maximum symmetrical spacetime. Furthermore,
 the difficulties in Ref.  \cite{tian} are investigated   in our work by  showing that the physical background of the topological charge is the sigma model (either linear or non-linear one,
 depending on the topological structure of the spacetime), and   this topological charge will be reconstructed in a more reasonable way inspired by Ref. \cite{kaku}.

 This paper is constructed as follows: in section 2, we will show our proposal to derive the global monopole spacetime exactly; in section 3, this proposal will be generalized to more general cases; we will discuss the thermodynamics of the global monopole spacetime in section 4 and a brief conclusion will be presented
 in section 5.

\section{The topological charge as the non-linear sigma model}
In this section, we show that the global monopole spacetime can be derived as an exact solution of Einstein equations if one applies the matter field as the non-linear sigma model. This section contains three parts, in part 1,  the matter field is introduced; in part 2, a derivation of the global monopole spacetime
 is presented; in part 3, the general properties of the Lagrangian which could be applied to derive the global monopole spacetime are discussed.
\subsection{The matter field}
As discussed before, the matter field is considered as the non-linear sigma model and whose Lagrangian is
\begin{equation}\label{non1}
L=-\frac{1}{2}\int(2-\frac{\phi^b\phi^b}{\eta^2})\nabla_{\mu}\phi^a\nabla^{\mu}\phi^a,
\end{equation}
where $\eta$ is a constant and the set $\phi^a, a=1, 2, 3...N$ is a set of $N$ real scalar fields. Also, the Latin letter "$a$" should be distinguished from the Greek letter "$\mu$", as the former is the inner index of the scalar fields and the  later is the tensor index. It should be also  noticed that the Einstein's notation is applied  for the Latin letters here.

  Applying the variational principle
  \begin{equation}
  \delta S=0,
  \end{equation}
 then the EOM of the scalar field is
  \begin{equation}\label{EOM}
  \frac{\nabla_{\mu}(\phi^b\phi^b)\cdot\nabla^{\mu}\phi^a}{\eta^2}+ (2-\frac{ \phi^b\phi^b}{\eta^2})\cdot\nabla_{\mu}\nabla^{\mu}\phi^a=-\frac{(\nabla_{\mu}\phi^b\nabla^{\mu}\phi^b)\phi^a}{\eta^2}.
  \end{equation}
The stress-energy tensor of the scalar field can be determined as
  \begin{equation}\label{td}
  T_{\mu\nu}=-\frac{2}{\sqrt{|g|}}\frac{\delta S_M}{\delta g^{\mu\nu}},
  \end{equation}
  and the result is
  \begin{equation}\label{non}
  T_{\mu\nu}=-\frac{g_{\mu\nu}}{2}(2-\frac{\phi^b\phi^b}{\eta^2})\nabla_{\lambda}\phi^a\nabla^{\lambda}\phi^a+(2-\frac{\phi^b\phi^b}{\eta^2})\nabla_{\mu}\phi^a\nabla_{\nu}\phi^a.
  \end{equation}
  \subsection{Derivation of the global monopole spacetime}
  The metric ansatz we apply here is
  \begin{equation}
  ds^2=-f(r)dt^2+h(r)dr^2+r^2d\theta^2+r^2\sin^2\!\theta d\varphi^2,
  \end{equation}
  where $f(r)$ and $h(r)$ are both just functions of the  variable $r$. In such ansatz, the non-vanishing components of the Einstein tensor are
  \begin{eqnarray}
  G_{00}&=&\frac{f(r)(h(r)^2+rh(r)'-h(r))}{h(r)^2r^2},\nonumber\\
    G_{11}&=&\frac{f(r)+rf(r)'-f(r)h(r)}{f(r)r^2},\nonumber\\
     G_{22}&=&-\frac{r\{h(r)rf(r)'^2+2h(r)^2f(r)^2\}}{4f(r)^2h(r)^2}\nonumber\\
     &-&\frac{rf(r)[rf(r)'h(r)'-2h(r)(f(r)'+rf(r)'')]\}}{4f(r)^2h(r)^2},\nonumber\\
         G_{33}&=&-\frac{r\sin^2\!\theta(h(r)rf(r)'^2+2h(r)^2f(r)^2)}{4f(r)^2h(r)^2}\nonumber\\
         &-&\frac{f(r)[rf(r)'h(r)'-2h(r)(f(r)'+rf(r)'')]}{4f(r)^2h(r)^2}.
  \end{eqnarray}
  By applying the Einstein equations
  \begin{equation}\label{ein}
  G_{\mu\nu}=8\pi G_nT_{\mu\nu},
  \end{equation}
  and the EOM of the scalar field $\phi^a$, one can find the solutions are
  \begin{eqnarray}
  f(r)=h(r)^{-1}=1-8\pi G\eta^2-\frac{2MG}{r},\nonumber\\
  \phi^1=\eta\sin\theta\cos\phi,  \phi^2=\eta\sin\theta\sin\phi, \phi^3=\eta\cos\theta,
   \end{eqnarray}
   so the geometrical structure  of the spacetime is obtained as
     \begin{eqnarray}\label{metric}
  ds^2=&-&(1-8\pi G\eta^2-\frac{2MG}{r})dt^2+\frac{1}{1-8\pi G\eta^2-\frac{2MG}{r}}dr^2\nonumber\\
  &+&r^2d\theta^2+r^2\sin^2\!\theta d\varphi^2.
  \end{eqnarray}
Thus the global monopole solution is derived strictly.

  Furthermore, inspired by Ref. \cite{kaku} the topological charge of this spacetime can be constructed as
  \begin{equation}
\tilde{Q}=\frac{1}{8\pi}\int \epsilon_{a_1a_2a_3}\phi^{a_1} d(\phi^{a_2})\wedge d(\phi^{a_3}).
\end{equation}
    Following such definition, the topological charge of the global monopole spacetime can be calculated as $\eta^3$.
    \subsection{Lagrangian that can be used to derive the global spacetime}

    In this subsection, we would like to discuss a class of Lagrangian which could be used to derive the global monopole spacetime. To begin with, we choose the ansatz for the Lagrangian as following
      \begin{equation}
      L=-\frac
      {1}{2}F(\phi^{b}\phi^{b})\nabla_{\mu}\phi^{a}\nabla^{\mu}\phi^{a}-V(\phi^{a}\phi^{a}).
  \end{equation}
  Then one can obtain the energy-stress tensor as
    \begin{equation}
     T_{\mu\nu}=F\nabla_{\mu}\phi^{a}\nabla_{\nu}\phi^{a}-g_{\mu\nu}(\frac{1}{2}F(\nabla_{\lambda}\phi^{a}\nabla^{\lambda}\phi^{a})+V(\phi^{a}\phi^{a})),
  \end{equation}
  and the equation of motion (EOM) of  the scalar field is
  \begin{equation}\label{eomxx}
     F\nabla_{\mu}\nabla^{\mu}\phi^{a}+F'\nabla_{\mu}(\phi^{b}\phi^{b})\nabla^{\mu}\phi^{a}=2V'\phi^{a}+F'(\nabla_{\mu}\phi^{b}\nabla^{\mu}\phi^{b})\phi^{a} ,
  \end{equation}
  in which $F'$ donates that the functional $F$ is varied by $\phi^a\phi^a$.

  By following \{Ref. \cite{Barriola}, we propose the ansatz of the scalar $\phi^a$ as
   \begin{equation}
    \phi^{a}=\eta g(r)\frac{x^{a}}{r},
  \end{equation}
    where $g(r)$ is a function of $r$. In their paper, the authors required that $g(r)\approx1$ while $r\to\infty$. However, in our discussion here, we would like to find the conditions that $g(r)$ could be equal to $1$ exactly,
   so the expression of the scalar field could be obtained as following
     \begin{eqnarray}\label{scalar}
    \phi^{1}&=&\eta \sin\theta\cos\varphi, \nonumber\\
    \phi^{2}&=&\eta \sin\theta\sin\varphi ,\nonumber\\
    \phi^{3}&=&\eta \cos\theta .
  \end{eqnarray}
   To attain such aim, we should first consider the global monopole spacetime, namely eq. (\ref{metric}). In such metric, the non-vanishing components of the Einstein tensor could be calculated as following,
\begin{eqnarray}
    G_{00}&=&(1-8\pi G\eta^2-\frac{2GM}{r})\frac{8\pi G\eta^2}{r^2},\nonumber\\
    G_{11}&=&-(\frac{8\pi G\eta^2}{r^2-8\pi G\eta^2 r^2-2GMr}).
    \end{eqnarray}
    Besides, the non-vanishing components of the energy -stress tensor are
    \begin{eqnarray}
      T_{00}&=&-\frac{(V(\eta^2)r^2+2F(\eta^2)\eta^2)(2GM+8\pi G\eta^2r-r)}{2r^3},\nonumber\\
      T_{11}&=&\frac{V(\eta^2)r^2+2F(\eta^2)\eta^2}{4GMr-2r^2+16G\pi\eta^2 r^2},\nonumber\\
      T_{22}&=&-\frac{V(\eta^2)r^2}{2},\nonumber\\
      T_{33}&=&-\frac{1}{2}V(\eta^2)r^2\sin^2\theta.
    \end{eqnarray}
    By comparing the equations together, one can find that once the conditions listed as the following was satisfied,  the Einstein equations could be verified,
    \begin{equation}\label{c1}
 V(\eta^2)=0, F(\eta^2)=1.
    \end{equation}
    Furthermore, the scalar field should satisfy to the EOM (\ref{eomxx}). By applying the results of Eq. (\ref{scalar}) to Eq. (\ref{eomxx}), one can obtain the EOM as
    \begin{equation}
    -\frac{2}{r^2}\phi^{a}=2V'(\eta^2)\phi^{a}+F'(\eta^2)\frac{2\eta^2}{r^2}\phi^{a}.
    \end{equation}
    As discussed above, we have $V(\eta^2)=0,$, so $V(\eta^2)'$ must be independent of the variable $r$. So one can obtain that,
    \begin{equation}\label{c2}
    V'(\eta^2)=0\;,\;F'(\eta^2)=-\frac{1}{\eta^2}.
    \end{equation}
    To conclude this section, we claim that once the non-linear model satisfies to the conditions listed in eqs. (\ref{c1}), (\ref{c2}), it could be used to derived the global monopole spacetime.
  \section{The generalization of the derivation}
In this section, the method proposed above is generalized to more general cases. Firstly, it will be applied to higher dimensional spacetime, and secondly, other classes of symmetrical spacetime with the global monopole charge will be
derived.
\subsection{The spherically symmetric case}
In this case, the metric ansatz of the $n$ dimensional spacetime is
\begin{equation}\label{ansatz}
ds^2=-f(r)dt^2+h(r)dr^2+r^2d\Omega_{n-2},
\end{equation}
where $\Omega_{n-2}$ represents the $n-2$ dimensional sphere. With some complicated calculations, one can verify that the non-vanishing components of Einstein tensor are
  \begin{eqnarray}
  G_{00}&=&\frac{f(r)[(n-2)(n-3)h(r)^2+(n-2)rh(r)']}{2h(r)^2r^2}\nonumber\\
    &-&\frac{(n-2)(n-3)f(r)}{2h(r)r^2},\nonumber\\
    G_{11}&=&\frac{(n-2)[(n-3)f(r)+rf(r)'-(n-3)f(r)h(r)]}{2f(r)r^2},\nonumber\\
     G_{ii}&=&\frac{g_{ii}}{4f(r)^2h(r)^2r^2}\{f(r)r[2h(r)((n-3)f(r)'+rf(r)'')\nonumber\\
     &-&rf(r)'h(r)']-h(r)[f(r)']^2r^2\nonumber\\
    &-&  2(n-3)(n-4)f(r)^2[h(r)^2-h(r)]\nonumber\\
    &-& 2(n-3)f(r)^2rh(r)'\},
  \end{eqnarray}
  where $g_{ii}$ is the components of the reduce metric of the $n-2$ dimensional sphere.

Besides, the matter field applied here is still the non-linear scalar field, meaning that Eq.(\ref{non}) can be also applied here. Combining the EOM of the scalar field (\ref{EOM}) and the Einstein equations (\ref{ein}), one can solve
the equations as
\begin{eqnarray}
ds^2=&-&(1-\frac{2G_nM}{r^{n-3}}-\frac{8\pi G_n\eta^2}{n-3})dt^2\nonumber\\
&+&\frac{1}{1-\frac{2G_nM}{r^{n-3}}-\frac{8\pi G_n\eta^2}{n-3}}dr^2+r^2d\Omega_{n-2},
\end{eqnarray}
and
\begin{eqnarray}
\phi^1&=&\eta \cos\theta_1,\nonumber\\
\phi^2&=&\eta \sin\theta_1\cos\theta_2,\nonumber\\
&\cdot&\nonumber\\
&\cdot&\nonumber\\
&\cdot&\nonumber\\
\phi^{n-2}&=&\eta \sin\theta_1\sin\theta_2...\sin\theta_{n-2}\cos\theta_{n-1},\nonumber\\
\phi^{n-1}&=&\eta \sin\theta_1\sin\theta_2...\sin\theta_{n-2}\sin\theta_{n-1}.
\end{eqnarray}
Then the topological charge should be defined as
\begin{equation}
\tilde{Q}=\frac{1}{(n-2)!A_{n-2}}\int \epsilon_{a_1a_2...a_{n-1}}\phi^{a_1} d(\phi^{a_2})\wedge...\wedge d(\phi^{a_{n-1}}),
\end{equation}
where $A_{n-2}$ is the volume of the $n-2$ dimensional sphere in $n$ dimensional spacetime
\begin{equation}
A_{n-2}=\pi^{\frac{n-1}{2}}\frac{\Gamma(\frac{n}{2})\Gamma(\frac{n-1}{2})...\Gamma(\frac{3}{2})}{\Gamma(\frac{n+2}{2})\Gamma(\frac{n+1}{2})...\Gamma(\frac{5}{2})},
\end{equation}
one can verify that the topological charge in the $n$ dimensional global spacetime  is $\eta^{n-1}$ by definition.

\subsection{The plane-symmetric case}
We derive the global monopole-like spacetime in the plane-symmetric case in this subsection. Firstly, we introduce  the metric ansatz as
\begin{equation}
ds^2=-f(r)dt^2+h(r)dr^2+r^2dx^idx_i.
\end{equation}
Here  Einstein's notation is applied  and the index $i$ range   is $i=2,3,4....n$. One can calculate the non-vanishing components of the Einstein tensor as following
\begin{eqnarray}
G_{00}&=&-\frac{(n-2)f(r)[(n-3)h(r)-rh(r)')]}{2h(r)^2r^2},\nonumber\\
G_{11}&=&\frac{(n-2)f(r)[(n-3)f(r)+rf(r)')]}{2f(r)r^2},\nonumber\\
G_{ii}&=&g_{ii}\frac{f(r)r\{2h(r)[(n-3)f(r)'+f(r)'']-rf(r)'h(r)'\}}{4f(r)^2h(r)^2r^2}\nonumber\\
&+&g_{ii}\frac{2(n-3)f(r)^2[(n-4)h(r)-rh(r)']}{4f(r)^2h(r)^2r^2}\nonumber\\
&-&g_{ii}\frac{h(r)r^2[f(r)']^2}{4f(r)^2h(r)^2r^2}.
\end{eqnarray}
Besides, the matter field applied here is the linear sigma model, which is different from that in the spherically symmetric case, and its Lagrangian is
\begin{equation}
L=\frac{1}{16\pi G}\int R-\frac{1}{2\eta^2}\int\nabla_{\mu}\phi^a\nabla^{\mu}\phi^a.
\end{equation}
By applying the variational principle again, the EOM of the scalar field can be obtained directly as
   \begin{equation}\label{EOMP}
\nabla_{\mu}\nabla^{\mu}\phi^a=0,
  \end{equation}
and the  stress-energy tensor can be calculated as the following by using Eq.(\ref{td}) again
    \begin{equation}
  T_{\mu\nu}=g_{\mu\nu}(-\frac{1}{2}\nabla_{\lambda}\phi^a\nabla^{\lambda}\phi^a)+\nabla_{\mu}\phi^a\nabla_{\nu}\phi^a.
  \end{equation}
  So the general solution of the Einstein equations in this case is
  \begin{eqnarray}
ds^2=&-&(-\frac{2 G_nM}{r^{n-3}}-\frac{8\pi G_n\eta^2}{n-3})dt^2\nonumber\\
&+&\frac{1}{-\frac{2 G_nM}{r^{n-3}}-\frac{8\pi G_n\eta^2}{n-3}}dr^2+r^2dx^idx_i,
\end{eqnarray}
and the solutions of the scalar fields are
\begin{eqnarray}
\phi^1&=&\eta x_2,\nonumber\\
\phi^2&=&\eta x_3,\nonumber\\
&\cdot&\nonumber\\
&\cdot&\nonumber\\
&\cdot&\nonumber\\
\phi^{n-2}&=&\eta x_{n-1}.
\end{eqnarray}
The topological charge is constructed as
\begin{equation}
\tilde{Q}=\frac{1}{(n-2)!A_{n-2}}\int \epsilon_{a_1a_2...a_{n-2}}d(\phi^{a_1})\wedge...\wedge d(\phi^{a_{n-2}}),
\end{equation}
where $A_{n-2}$ is the volume of a unit submanifold and the topological charge here is $\eta^{n-2}$. We should mention that the plane-symmetric space is not compact, so in order to make the concept of 'topological charge' more rational, the technology of compactification should be applied here, and one can refer to Ref. \cite{klem,vanzo} for more details. In this case, our result can match well with the topological charge proposed in Ref. \cite{tian}. Thus, the proposal of Ref. \cite{tian} is
just one  of our possible situation. Moreover, our calculation reveals that the physical origin of the topological charge is related to the global monopole charge, namely the sigma model
exactly.
\subsection{The hyperbolically symmetric case}
We apply our proposal to hyperbolically symmetric case in this subsection, and the metric ansatz is
\begin{equation}\label{ansatz}
ds^2=-f(r)dt^2+h(r)dr^2+r^2d\theta^2+r^2\sinh\!\theta^2\Omega_{n-3},
\end{equation}
with calculation, the non-vanishing components of the Einstein tensor are
 \begin{eqnarray}
  G_{00}&=&\frac{f(r)[(n-2)rh(r)'-(n-2)(n-3)h(r)^2}{2h(r)^2r^2}\nonumber\\
  &-&\frac{(n-2)(n-3)h(r)]}{2h(r)^2r^2},\nonumber\\
    G_{11}&=&\frac{(n-2)[(n-3)f(r)+rf(r)'+(n-3)f(r)h(r)]}{2f(r)r^2},\nonumber\\
   G_{ii}&=&\frac{g_{ii}}{4f(r)^2h(r)^2r^2}\{f(r)r[2h(r)((n-3)f(r)'+rf(r)'')\nonumber\\
     &-&rf(r)'h(r)']-h(r)[f(r)']^2r^2\nonumber\\
    &+&  2(n-3)(n-4)f(r)^2[h(r)^2+h(r)]\nonumber\\
    &-& 2(n-3)f(r)^2rh(r)'\} .
  \end{eqnarray}
We argue that the matter field here is also the non-linear sigma model, which means that the Lagrangian (\ref{non1}) can still be applied in this case. However, it should be underscored that the space where the scalar fields
live in is Minkowski space, namely the EOM of the scalar fields is
 \begin{eqnarray}\label{EOMH}
& &\eta_{bd} \nabla_{\mu}(\phi^b\phi^d)\cdot\nabla^{\mu}\phi^a+  \eta^2(2-\eta_{bd}\frac{ \phi^b\phi^d}{\eta^2})\cdot\nabla_{\mu}\nabla^{\mu}\phi^a\nonumber\\
 &=&-(\eta_{bd}\nabla_{\mu}\phi^b\nabla^{\mu}\phi^d)\phi^a,
  \end{eqnarray}
  and the stress-energy tensor is
  \begin{eqnarray}
  T_{\mu\nu}=-&\frac{g_{\mu\nu}}{2}&\eta_{ac}(2-\eta_{bd}\frac{\phi^b\phi^d}{\eta^2})\nabla_{\lambda}\phi^a\nabla^{\lambda}\phi^c+\nonumber\\
 &\eta_{ac}& \eta_{ac}(2-\eta_{bd}\frac{\phi^b\phi^d}{\eta^2})\nabla_{\mu}\phi^a\nabla_{\nu}\phi^c.
  \end{eqnarray}

  Using the Einstein equations (\ref{ein}) and the EOM of the scalar fields  (\ref{EOMH}), one can obtain the spacetime structure
  \begin{eqnarray}\label{ansatz}
ds^2=&-&(-1-\frac{2 G_nM}{r^{n-3}}-\frac{8\pi G_n\eta^2}{n-3})dt^2\nonumber\\
&+&\frac{1}{-1-\frac{2 G_nM}{r^{n-3}}-\frac{8\pi G_n\eta^2}{n-3}}dr^2+r^2d\theta^2\nonumber\nonumber\\
&+&r^2\sinh\!\theta^2\Omega_{n-3},
\end{eqnarray}
and the solutions of the matter fields
\begin{eqnarray}
\phi^1&=&\eta \cosh\theta_1,\nonumber\\
\phi^2&=&\eta \sinh\theta_1\cos\theta_2,\nonumber\\
&\cdot&\nonumber\\
&\cdot&\nonumber\\
&\cdot&\nonumber\\
\phi^{n-2}&=&\eta \sinh\theta_1\sin\theta_2...\sin\theta_{n-2}\cos\theta_{n-1},\nonumber\\
\phi^{n-1}&=&\eta \sinh\theta_1\sin\theta_2...\sin\theta_{n-2}\sin\theta_{n-1}.
\end{eqnarray}
Indeed, one can construct the topological charge density in this case as
\begin{equation}
Q=\frac{1}{(n-2)!A_{n-2}}\int \epsilon_{a_1a_2...a_{n-1}}\phi^{a_1} d(\phi^{a_2})\wedge...\wedge d(\phi^{a_{n-1}}),
\end{equation}
and the result is $\eta^{n-1}$, and just like plane-symmetric case,  compactification should be also applied here.
\section{The thermodynamical properties of the global monopole spacetime}

In Ref. \cite{tian}, while the authors  constructed the black hole thermodynamics with the topological charge in formula, it is interesting to investigate the matter field in the spacetime that play an important role in the thermodynamics. In our proposal, we suggest that the matter
field is the sigma model, and for convenience, we propose the first law of thermodynamics in the plane case as
\begin{equation}
dE=TdS-\frac{16\pi G_n\tilde{Q}^\frac{4-n}{n-2}S'}{A_{n-2}(n-2)^2(n-3)}d\tilde{Q},
\end{equation}
in which $S$ is the entropy of the black hole,  $S=\frac{V_{n-2}r^{n-2}}{4G_n}$, and $S'$ denotes that the entropy $S$ is derived by the variable $r$. One point should be underscored here is that the power of the topological charge in our proposal of the first law of the thermodynamics is the same as that in the propose in Ref. \cite{tian}, and the coefficient is chosen for convenience, implicating that our propose is reasonable. With some calculation, one can get the following equation
\begin{equation}\label{therm}
dM+\frac{4\pi\tilde{Q}^\frac{4-n}{n-2}r^{n-3}|_{r_h}}{(n-2)(n-3)}d\tilde{Q}=TdS.
\end{equation}
    Here, it should be noted that there are two meanings of the symbol $"d"$, namely if it acts on the variable of the spacetime, it stands for the normal differential, whereas if it acts on the global parameters, it means variation.

Indeed, there is a  quasi-local version of such black hole thermodynamics, which is the so-called unified first law,
\begin{equation}
(dE)_a=A_{n-2}(\psi)_a+w(dV_{n-1})_a,
\end{equation}
which was proposed in Ref. \cite{hayward}, and $V_{n-1}=\frac{A_{n-2}r^{n-1}}{n-1}$. Here $w$ is the work term, defined as the trace of the stress-energy in leading two dimensions, namely
\begin{equation}\label{work}
w=-\frac{4\pi}{(n-2)A_{n-2}}(g^{tt}T_{tt}+g^{rr}T_{rr}),
\end{equation}
and $\psi_a$ is known as the energy flux,
\begin{equation}
\psi^a=T^{ab}\nabla_br+w\nabla^ar.
\end{equation}

  In the quasi-local version of thermodynamics, the quasi-local energy $E$ can  be defined as the so-called Misner-Sharp energy
\begin{equation}\label{misner}
E_{ms}=\frac{r^{n-3}}{2G_n}(k-g^{rr}),
\end{equation}
see Ref. \cite{misner,Maeda,zhang201}. In our discussion, the value is
\begin{equation}
E_{ms}=M+\frac{4\pi\tilde{Q}^\frac{2}{n-2}}{n-3}r^{n-3},
\end{equation}
varying it, one has
\begin{equation}
d E_{ms}=d M+\frac{4\pi\tilde{Q}^\frac{4-n}{n-2}r^{n-3}}{(n-2)(n-3)}d\tilde{Q}+4\pi\tilde{Q}^\frac{2}{n-2}r^{n-4}dr,
\end{equation}
comparing this with Eq.(\ref{therm}), we can get the unified first law in this case as
\begin{equation}\label{uni1}
dE_{ms}=TdS+4\pi\tilde{Q}^\frac{2}{n-2}r^{n-4}dr,
\end{equation}
it should be noted that the global monopole charge plays a significant role in the quasi-local version of the spacetime thermodynamics.

Besides, there are a series of works which generate the spacetime construction from thermodynamics by using the unified law and treating the spacetime as an adiabatic system Ref. \cite{zhang,zhang1,zhang2}. These works can be summarized as follow:  one can derive the structure of the spacetime once  there is a work term  defined by Eq.(\ref{work}). However, there is a great limitation in these works: the authors
only applied this method to the spacetime with the maximum symmetry. So the problem is raised naturally: could such method be applied to more general situations?

Indeed, there was a try before. In Ref. \cite{tan}, the authors put forward a
similar method by using the Komar mass and the ADM mass, rather than the Misner-Sharp mass to generate spacetime from thermodynamics. In that work, the authors applied their proposal to the global monopole spacetime successfully. However, when they generated  the global monopole spacetime
some subtle trick was applied, which may cause argument.

Here, we argue that the original method can be applied to the spacetime with the global charge exactly. We consider the plane case firstly and choose the metric ansatz for convenience as
\begin{equation}
ds^2=-f(r)dt^2+\frac{1}{h(r)}dr^2+r^2dx_idx^i.
\end{equation}
In this case, the Misner-Sharp energy can be calculated as
\begin{equation}
E_{ms}=-\frac{r^{n-3}}{2G_n}h(r),
\end{equation}
we can use the unified first law in a adiabatic condition to derive $h(r)$. According to Eq. (\ref{uni1}), we can get the equation of $h(r)$ as follow,
\begin{equation}
-\frac{1}{2G_{n}}[(n-3)h(r)r^{n-4}+h(r)'r^{n-3}]dr=4\pi\tilde{Q}^\frac{2}{n-2}r^{n-4}dr.
\end{equation}
The solution of this differential equation is
\begin{equation}
h(r)=-\frac{2 G_nM}{r^{n-3}}-\frac{8\pi G_n\eta^2}{n-3}.
\end{equation}
Here "M" is the integration constant, known as the ADM mass.

Our next task is to find $f(r)$. To do this we need to apply the so-called geometric surface gravity, which was proposed in Ref. \cite{hayward} firstly and generalized to higher dimensional spacetime in Ref. \cite{zhang1},  defined as follow
\begin{equation}
\kappa_G=(n-3)G_n\frac{E_{ms}}{r^{n-2}}- A_{n-2}G_nrw.
\end{equation}
In the topic we discuss here, the work term $w$ is
\begin{equation}
w=\frac{4\pi}{A_{n-2}}\frac{\eta^2}{r^2}.
\end{equation}
So the geometric surface gravity can be calculated as
\begin{equation}
\kappa_G=\frac{(n-3)G_nM}{r^{n-2}}.
\end{equation}
The traditional surface gravity is defined as
\begin{equation}
\kappa= \frac{h(r)}{2f(r)}f(r)',
\end{equation}
by following the previous works, we assume that the surface gravity is equal to the geometric surface gravity, thus
\begin{equation}
\frac{h(r)}{2f(r)}f(r)'=\frac{(n-3)G_nM}{r^{n-2}},
\end{equation}
by solving the equation, we have
\begin{equation}
f(r)=-\frac{2 G_nM}{r^{n-3}}-\frac{8\pi G_n\eta^2}{n-3}.
\end{equation}
So the metric can be derived as
\begin{eqnarray}
ds^2&=&-(-\frac{2 G_nM}{r^{n-3}}-\frac{8\pi G_n\eta^2}{n-3})dt^2\nonumber\\
&+&\frac{1}{-\frac{2 G_nM}{r^{n-3}}-\frac{8\pi G_n\eta^2}{n-3}}dr^2+r^2dx_idx^i.
\end{eqnarray}
Furthermore, the previous discussion can be applied to the global monopole spacetime. However, in this situation, the thermodynamics of the spacetime with topological charge should be redefined as
\begin{equation}\label{therm1}
dM+\frac{4\pi\tilde{Q}^\frac{3-n}{n-1}r^{n-3}|_{r_h}}{(n-1)(n-3)}d\tilde{Q}=TdS.
\end{equation}
On the other hand, the Misner-Sharp energy is
\begin{equation}
E_{ms}=M+\frac{4\pi\tilde{Q}^\frac{2}{n-1}}{n-3}r^{n-3},
\end{equation}
and the unified first law is
\begin{equation}
dE_{ms}=TdS+4\pi\tilde{Q}^\frac{2}{n-1}r^{n-4}dr.
\end{equation}
Using the same method proposed above, one can get the global monopole solution
\begin{eqnarray}
ds^2=&-&(1-\frac{2 G_nM}{r^{n-3}}-\frac{8\pi G_n\eta^2}{n-3})dt^2\nonumber\\
&+&\frac{1}{1-\frac{2 G_nM}{r^{n-3}}-\frac{8\pi G_n\eta^2}{n-3}}dr^2+r^2d\Omega_{n-2}.
\end{eqnarray}
The same discussion can be applied to the hyperbolic symmetrical case directly.

Here, we generalize the method that one can derive the spacetime structure from thermodynamics to the spacetime with global monopole charge, which is not the maximum symmetrical spacetime, and the topological charge plays a centre role in the work
term.
\section{Discussion and Conclusion}
The global monopole spacetime was first presented in     Ref. \cite{Barriola}, in which the matter field is treated as an approximate one. Thus, one of our works is  deriving this spacetime exactly and revealing  that
the matter field which can be used to derive this spacetime is the non-linear sigma model. Moreover, the derivation is also generalized to higher dimensional spacetime ($n>4$) and the spacetime with other class of
symmetry, namely the plane symmetry and the hyperbolic symmetry. The plane-symmetric case was proposed in Ref. \cite{chen}, where the matter field is applied as a
charged complex scalar field. Nevertheless, in our work we find that the linear sigma model can also be used to derive the same spacetime structure. Exactly, a similar idea which argues that the matter field of the global monopole spacetime is a sigma model was first proposed in Ref. \cite{gib}. However, it should be mentioned that in such paper, the author just gave us a very simple claim without any detail in such discussion. Moreover, in their proposal, the authors
only focused on the spherically symmetric case in four dimensional spacetime. Thus we believe that our work makes an improvement  in such aspect of research.

Our another work is that we reconstruct the topological charge which was first proposed in Ref. \cite{tian}.  We find that the physical origin of the topological charge is the sigma model (both the linear one and the non-linear one). Moreover, our result can match the original proposal well in  the plane-symmetric case,  so it could be  claimed  that our work is an apposite generalization  of the construction
in Ref. \cite{tian}.

Moreover, we investigate the thermodynamics of the spacetime by applying Misner-Sharp energy in our discussion, and then we construct the unified first law for this class of spacetime. Additional, we propose that the global monopole spacetime can be generated from the unified first law.
\section{Acknowledgments}
This research is supported by the National Natural Science Foundation of China under Grant Nos. 11273009 and 11303006.

\end{CJK*}
\end{document}